\documentclass[twocolumn,twoside,slac_two]{revtex4}
\usepackage{graphicx}
\usepackage{fancyhdr}
\begin{document}

\title{Statistical Analysis of Multiwavelength Light curves}

%

\author{S. Larsson}
\affiliation{Department of Astronomy and Department of Physics, 
Stockholm University, AlbaNova, SE-10691 Stockholm, Sweden, 
The Oskar Klein Centre for Cosmoparticle Physics, 
Stockholm, Sweden}

\begin{abstract}
Since its launch in 2008 the Fermi Large Area Telescope provides
regular monitoring of a large sample of gamma-ray sources
on time scales from hours to years. Together with
observations at other wavelengths it is now possible
to study variability and correlation properties in a much more
systematic and detailed way than ever before. 
The paper describe some of the statistical methods and tools that
have been, or can be, used to characterize variability and
to study the relation between multiwavelength light curves.
Effects and limitations due to time sampling, 
measurement noise, non-stationarity etc are illustrated 
and discussed.

\end{abstract}

\maketitle

\thispagestyle{fancy}


\section{INTRODUCTION}

The combination of gamma-ray and radio observations together with
measurements in other wavelength bands has created extraordinary
opportunities to study multiwavelength properties of Active Galactic 
Nuclei, in particular for blazars. Among the aims are to better answer 
questions like,
how are the components of the spectral energy distribution (SED) 
related? Do the components originate from one or more spatial 
regions? Does Compton seed photons originate locally or from
a source external to the jet? Are hadronic cascades an important
contribution to the gamma-ray emission?
There are theoretical predictions of potentially observable effects 
that can be searched for, such as spectral softening associated 
with particle cooling during flare decay~\cite{dotson2011}.
From an observational point of view however, the first aim is to 
characterize source variability and correlations between different
spectral bands. Simultaneously the wealth of new data also allow us
to search for new and previously unobserved phenomena. 

The increasing amount of multiwavelength data on blazars
allow us to attack some of the unresolved questions in a
statistically more comprehensive way instead of studies based on
a restricted number of individual cases. 
The importance of this is particularly clear when one considers
the complex and apparently unsystematic multiwavelength
behaviour revealed by earlier investigations.

Available data from different spectral bands differ in terms
of Signal-to-Noise, time sampling, observation length etc. The 
interpretation of observed variability and correlations is
complicated not only by measurment noise and sampling but also
by the stochastic variability itself, which means that
observed variability properties may change with time and that
even unrelated time series can exhibit chance correlations
in time limited observations. All these effects needs to be considered 
in the analysis and interpretation of the observations.

\subsection{Data Properties}

Fermi Large Area Telescope (LAT) has a large field of view, covering 
about 20 $\%$ of the sky
at any particular time. Except for some shorter pointed observations
it is operated in a sky surveying mode, mapping the full sky every 3 hours.
This provides a regular monitoring of all the gamma-bright 
blazars on the sky. In total, about 1000 blazars were detected in 
the data of the first two years of operation and included in the
second LAT AGN catalog~\cite{2LAC}.
The signal-to-noise is such that a few tens of sources are typically
detected in time bins of one to a few days, while hundreds of AGNs
are significantly measured in time bins of one to a few weeks.

The regular sampling of the Fermi LAT observations is a great advantage
in the data analysis, not just for the gamma-ray analysis itself
but also for the analysis of observations at other wavelengths.
Radio and optical observations are often of high signal-to-noise
but the time sampling is in most cases less regular. It is then
an advantage that the sparse data can be compared with a
more densely covered light curve.

One limitation with Fermi light curves is however, that it is often
not possible to choose a single bin width that both resolve 
variability at high flux levels and at the same time detect 
the source at low flux levels. 
This commonly results in upper limits which are hard to handle
in the high level analysis. A way to remedy this is to use a
non-constant bin width chosen to give approximately the same
signal-to-noise for each bin. A procedure to create such
adaptive binned light curves for Fermi data is described in~\cite{lott2012}.
An alternative approach based on Bayesian Blocks has also
been developed~\cite{scargle2012}.

While Fermi has been operating for 3.5 years many blazars have
been followed in optical and radio for tens of years. This provides
valuable information on variability on longer time scales than
is accessible by Fermi. This has a bearing on for example duty cycles 
and on the question of non-stationarity of the variability.

\subsection{Analysis tools}

The analysis tools that have been applied to study blazar
variability include, excess variance, flare profile fitting,
flux distribution (duty cycles), power density spectra,
auto correlation function, structure function, and wavelets.
Multiwavelength correlations can be investigated by
e.g. direct light curve comparisons, flux - flux plots,
the cross correlation function and the cross spectrum. 
An overview of these methods with direct relevance to
blazar variability analysis is given in~\cite{scargle2011}.
Here we will instead focus on some of the practical aspects of cross
correlation analysis. 

\section{Cross Correlation}

For two discrete, evenly sampled light curves, $x(t_i)$ and $y(t_i)$, 
the Cross Correlation Function (CCF) as a function of 
time lag $\tau$ is,

\begin{equation}\label{eq:ccf}
CCF(\tau) = \frac{1}{N}\sum_{i=1}^{N} \frac{[x(t_i)-\bar{x}] 
[y(t_i \!-\! \tau)-\bar{y}]}{\sigma_x \sigma_y}
\end{equation}

\noindent
Where $\bar{x}$, $\sigma_x$ and $\bar{y}$, $\sigma_y$ are
mean and standard deviation for each of the two light curves.

The presence of measurement noise increases the standard deviation
in the data and hence reduces the correlation amplitudes as
computed by eq. 1. This bias can be removed by replacing
$\sigma_x \sigma_y$ in the equation by 
$\sqrt{(\sigma^2_x - e^2_x) (\sigma^2_y - e^2_y)}$,
where $e_x$ and $e_y$ are the measurement errors.

Long term astronomical light curves, such as those produced
by blazar monitoring programs, are almost always unevenly
sampled or contain data gaps. The two main approaches used
to handle uneven sampling in cross correlation
analysis is the Interpolated Cross Correlation Function (ICCF) 
and the Discrete Cross Correlation Function (DCCF).
These methods as well as the Z-transformed Discrete Correlation Function 
(ZDCF) are briefly described below.

\subsection{Recipes to calculate the Cross Correlation Function for 
Unevenly Sampled Light curves}

\subsubsection{The Discrete Cross Correlation Function, DCCF}

In the DCCF (or DCF) method, introduced by~\cite{edelsonkrolik},
a contribution to the DCCF is calculated only using the actual
data points. Each pair of points, one from each of the two
light curves, gives one correlation value at a lag corresponding
to their time separation. For two light curves with N and M
data points respectively this gives an Unbinned Cross Correlation
Function (UCCF),

\begin{equation}\label{eq:uccf}
UCCF_{ij} = \frac{(x_i-\bar{x})(y_j-\bar{y}) }{\sigma_x \sigma_y}
\end{equation}

The DCCF is then obtained by averaging the UCCF in time lag bins.
To illustrate this we use the two light curves of PKS~1510-089 
shown in Figure~1 to compute the UCCF and DCCF. These are
presented in Figure 2. 
It is in general recommended to calculate, for each DCCF lag, the 
means and variances using only the data points in the overlap 
interval~\cite{white1994} \cite{welsh1999}. This is sometimes
referred to as the {\it local DCCF}.

\begin{figure}
\centering
\includegraphics[width=75mm]{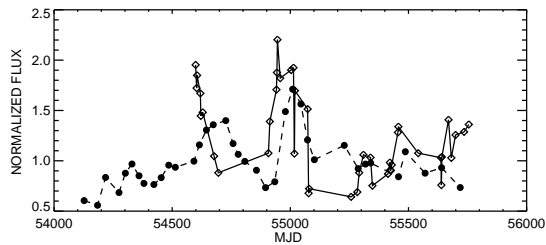}
\caption{Two light curves for the blazar PKS 1510-089.
The diamond symbols (solid curve) are observations at
sub-mm wavelength (0.87 mm) with the APEX telescope~\cite{larsson2012}
and the filled circles (dashed curve) are 
2 cm observations from the FGAMMA project~\cite{angelakis2008}.} 
\label{PKS1510lc}
\end{figure}

\begin{figure}
\centering
\hspace{-15mm}\includegraphics[width=95mm]{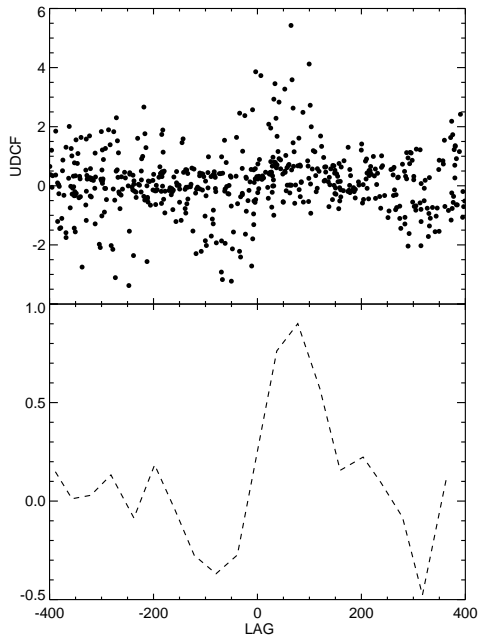}
\vspace{-15mm}
\caption{The unbinned (top) and binned (bottom) discrete cross correlation 
function for the two light curves in Figure 1.} \label{UDCFDCCFPKS1510}
\end{figure}

A point to notice is that a large variation in observational coverage
over the light curve can have a strong effect on the DCCF amplitudes.
This is particularly true if e.g. flares are densely covered with
observations while more quiescent levels are observed more
sparsely. The remedy is to rebin close data points to 
some minimum separation.

\subsubsection{The Interpolated Cross Correlation Function, ICCF}

In the ICCF method~\cite{gaskell1987}~\cite{gaskell1986},
the light curve is linearly interpolated and resampled onto a regular grid.
It is common to calculate the ICCF twice (where interpolation is
done in each of the light curves, one at a time) and average these.

\subsubsection{The Z-transformed Discrete Correlation Function, ZDCF}

The ZDCF method~\cite{alexander1997}, is base on the DCCF method 
but with a number of modifications.
The first is that DCCF bin widths are non-equal and chosen so that the
number of points averaged is the same for each bin. (This can
of course also be applied to the usual DCCF). The second is the
z-transformation of the DCCF. This transforms
the DCCF into a version for which the sampling distribution 
is closer to normal and allows for more robust estimation.  
Mean and variance are calculated for z and the resulting
DCCF with errors is then obtained by an inverse transformation of z.

\subsection{Differences between the CCF methods}

It has been shown that when the number of points is small, the ICCF 
method can have an advantage relative to the DCCF~\cite{white1994}. 
On the other hand the performance of the ICCF is dependent
on how well the interpolations between points can be made. When the
variability on time scales smaller than the point separation is
large the interpolation becomes unreliable. 
The ZDCF has been found to have an advantage over
the ICCF and DCCF, primarily for undersampled (relative to
the variability) light curves with few data points~\cite{alexander1997}.
For well sampled data however, all three methods produce
consistent results~\cite{smithvaughan2007}.

\subsection{Estimate of correlation strength and time lag}

The CCF does not preserve all information contained in the light curves.
It is useful mainly because it extract some aspects of the
relation between the two light curves, in particular the amplitude
and time lag of correlated variability. Therefore, the aim of the CCF
analysis is, in general, to estimated the significance of a correlation
peak, its amplitude and time lag with uncertainties.

\subsubsection{Model Dependent Monte Carlo Methods}

Uncertainties of computed CCFs can be estimated by comparison with
the analysis of simulated light 
curves~\cite{chatterjee2008}~\cite{maxmoerbeck2012}. 
The analysis consists of

\begin{itemize}
\item Monte Carlo simulation of synthetic light curves. The variability
is in general assumed to be described by red noise and is typically
simulated by the method decribed in~\cite{timmer1995}. The time
resolution should typically be higher than that of the observations
(if resampling is needed in step 2).
\item Sample the simulated time series at the same times and bin widths
as the observations. 
\item Compute the CCF for each pair of simulated light curves and 
determine the time lag and amplitude at the peak.
\item Repeat N times and compare the distribution in time lag and
amplitude with those of the data to estimate uncertainties.
\end{itemize}

\subsubsection{Model Independent Monte Carlo Methods}

A model independent Monte Carlo method to estimate uncertainties due 
to measurement  noise and uneven sampling was introduced 
by~\cite{peterson1998}. 
This is a recipe to compute uncertainties 
in a time lag determination, not in the individual DCCF values, 
since the latter are in general strongly correlated.
The error estimation proceeds through the following steps

\begin{itemize}
\item Add white noise with standard deviation equal to
the 1 sigma errors to the data.
\item Make a bootstrap-like selection of data points.
\item Compute the CCF and determine the time lag at the peak.
\item Repeat N times and compute rms(lag).
\end{itemize}

An illustration of the distribution of lags used to
estimate the uncertainty is shown in Figure 3. This
is known as the Cross Correlation Peak 
Distribution~\cite{maoznetzer1989}.
The method assumes that the variance in position of
the lag peak depends linearly on the variance of white noise
in the light curve. When the Signal-to-noise is low this is
no longer the case. One can then choose to add 
the three noise components (white noise in light curve 1, 2
and the bootstrap selection) one at a time and in a final
step add the three lag variances. 

In most cases the determination of the CCF peak position is best
done by fitting a function (e.g. a Gaussian) to the peak. The
range of the fit should be wide enough to consistently
find the peak in a large number of simulations but
not so wide that it is determined by the base~\cite{wadehorne1988}. 

\subsubsection{Mixed Source Correlation}
To estimate the significance of correlation peaks in the presence 
of stochastic variability, one approach that has been used is
based on simulated red noise light curves as described in section 2.3.1.
Since a fairly large number of radio and gamma-ray light curves 
of blazars is now available an alternative approach is possible.
One can study the probability of chance correlations by 
{\it mixed source correlation}. This is done 
by correlating a radio light curve for a source with the gamma-ray 
light curves of all the other sources and compare the
chance correlations in these DCCFs with any observed peak
in the actual DCCF for the source (see Figure 4). Under the assumption 
that the variability properties
are similar for the different sources, the probability
distribution of the resulting DCCFs
should reflect the occurrence of chance correlations. An important advantage of 
the method is that it does not require any characterization of the 
variability properties, except the assumption that these are similar
for all the sources used in the mixed analysis. Compared to
light curve simulations the disadvantage 
is that the number of light curves that can be used is limited.

The mixed source correlation method has been applied to the
correlation of Fermi and Fgamma radio light 
curves~\cite{fuhrmann2012}. In that case the time sampling of the
radio light curves differ from source to source, while all
gamma-ray light curves have the same time bins. An identical
relative time sampling is therefore achieved when a single
source radio light curve is correlated with the gamma-ray
light curves of all other sources with similar properties.

\begin{figure}
\centering
\hspace{15mm}\includegraphics[width=80mm]{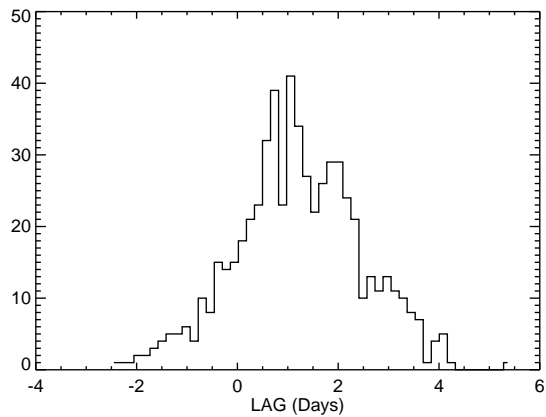}
\vspace{0mm}
\caption{Distribution of 600 lag estimates using the model
independent approach~\cite{peterson1998}. The two
input light curves contained sinusoidal oscillations
with zero intrinsic lag. Due to white noise the
actual lag determined from the correlation peak was
near +1.2 days. The method adds an additional white
noise component with the same rms as in the initial light curves.
Many such pairs are produced and the distribution of the
estimated lags is used to estimate an uncertainty for the lag.
} \label{lagdistrib}
\end{figure}

\begin{figure}
\centering
\hspace{-25mm}\includegraphics[width=95mm]{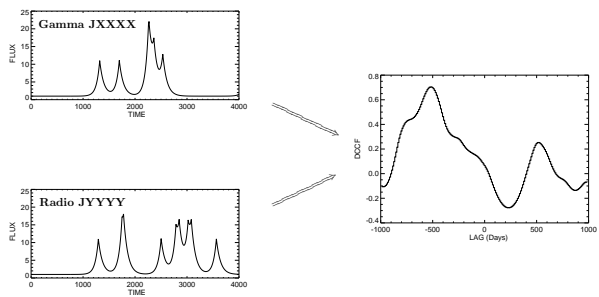}
\vspace{-65mm}
\caption{In mixed source correlation unrelated light curves are
correlated and the resulting DCCF is used to estimate the
probability of chance correlations.} \label{mscfig}
\end{figure}

\subsubsection{Peak significance estimate using Bartlett's formula}

As described in~\cite{smithvaughan2007} it is also possible
to estimate significances using Bartlett's formula~\cite{bartlett1955}. 
Under the null hypothesis, i.e. no intrinsic correlation, the variance 
of the distribution of CCF points are,
\begin{equation}\label{eq:bartlett}
\sigma^2(CCF(\tau_j)) = \frac{1}{N_{pair}}\sum_{i=1} ACF_1(\tau_i)ACF_2(\tau_i)
\end{equation}

where $ACF_1$ and $ACF_2$ are the autocorrelation functions of the 
two light curves and $N_{pair}$ the number of pairs contributing at
the lag $\tau_j$.

\section{Other issues}

\subsection{The effect of trends on the CCF}

If one of the two correlated light curves contains an unrelated
background trend the correlation function can be severely
distorted. In such cases it is necessary to detrend the
light curve, e.g. by polynomial subtraction, before calculating
the CCF. A detrending may increase or decrease the signal-to-noise
in the lag determination. This is because long time scales have few points 
and tend to be noisy but if detrending removes most of the signal 
we are left with noise.

It is also worth noting that if the light curves
contain correlated variability on different time scales it
is often useful to compute CCFs both with and without detrending,
since these can be sensitive to different variability time scales.

\subsection{Complex correlations}

Many investigations have been made of correlations between 
blazar light curves for different wavelength bands. Strong
correlations are often seen between optical and gamma-rays,
while the radio - optical and radio - gamma-ray correlations
appear to be weaker.
The SMARTS optical/near-IR monitoring of Fermi sources
has been used to study correlations with the
gamma-ray variability for six blazars~\cite{bonning2012}. 
Correlated variability were reported for these sources, except for
the high synchrotron peaked BL Lac, PKS 2155-304. Only one of the
sources, PKS 1510-089, showed evidence for a time lag. 
The same 2-week lag of optical relative
to gamma-rays that was seen from SMARTS data for this source 
was also described in~\cite{abdo_pks1510}
where it was further shown that each of the three analyzed
flares exhibited a similar time lag. A detailed comparison
of the light curves showed that the envelope (start and end) of
each flare is about the same in the two bands but the
shape of the flare is different. The ratio of gamma-ray
to optical flux is higher in the beginning of the flare
than towards the end. On top of this the flares exhibit
strong variability on shorter time scales which is partly
correlated and responsible for a correlation peak near zero lag. 

As shown in~\cite{maxmoerbeck2012} and~\cite{fuhrmann2012} the 
radio - gamma-ray correlation is in general weak. While
some gamma-ray flares appear to have a radio counterpart
others do not, even for the same object. Other examples
of variable correlation properties are flares that are seen
in only one band, {\it orphan flares}~\cite{abdo2010}, 
and significant changes in lag with
time~\cite{chatterjee2008}. It is clear that 
the multiwavelength correlation properties of blazars are
complex and that the studies up to now have not been able
to disentangle this complexity. This shows that it is not
sufficient to make intensive studies of individual sources
but these needs to be complemented with wider statistical 
analysis of a larger sample of sources with the aim to make 
a synthesis of the variability and correlation properties and 
to investigate dependencies on source type and other properties.

\subsection{Non-stationarity?}

From available observations in all spectral ranges it appears that
blazars exhibit variability on the longest
accessible time scales. In shorter observations we can therefore 
expect the variability to show non-stationarity,
such that e.g. mean and variance may differ from the long
term mean. Two ways to limit such effects
are to detrend the light curves and to use the local mean and
variance as described above and in~\cite{welsh1999}. 

For longer time series it is possible to study non-stationary
effects by dividing the data in segments and follow the 
evolution of the Power Density Spectrum and the CCF with time. 
A discussion of this topic can be found in~\cite{scargle2011}
and references therein. 


\bigskip 
\begin{acknowledgments}
The author likes to thank the colleagues in the Fermi collaboration
for inspiring and helpful discussions and is also grateful to the 
Swedish National Space Board for funding of this work.

\end{acknowledgments}

\bigskip 

\begin{thebibliography}{9}   
\bibitem{abdo2010}
Abdo, A. A., et al. 2010a, Nature, 463, 919
\bibitem{abdo_pks1510}
Abdo, A. A., et al. 2010b, Astrophys. J., 721, 1425
\bibitem{2LAC}
Ackermann, M., et al., 2011, Astrophys. J., 743, 171 
\bibitem{alexander1997}
Alexander T., 1997, in Maoz D., Sternberg A., Leibowitz E. M., 
eds, ASSL Vol. 218, Astronomical Time Series, 163
\bibitem{angelakis2008}
Angelakis, E., et al., 2008, Memorie della Societa Astronomica Italiana, 
79, 1042 
\bibitem{bartlett1955}
Bartlett M. S., 1955, An Introduction to Stochastic Processes,
CUP: Cambridge
\bibitem{bonning2012}
Bonning, E. W. et al., 2012, submitted to Astrophys. J., arXiv:1201.4380
\bibitem{chatterjee2008}
Chatterjee, R., et al., 2008, Astrophys. J., 689, 79 
\bibitem{dotson2011}
Dotson, A., Georganopoulos, M., Kazanas, D. \& Perlman, E., 2011,
2011 Fermi Symposium proceedings - eConf C110509, arXiv:1111.6551
\bibitem{edelsonkrolik}
Edelson, R. A., \& Krolik, J. H., 1988, Astrophys. J., 333, 646 
\bibitem{fuhrmann2012}
Fuhrmann, L. et al., 2012, in preparation
\bibitem{gaskell1987}
Gaskell M. C. \& Peterson B. M., 1987, Astrophys. J. Suppl, 65, 1
\bibitem{gaskell1986}
Gaskell, C. M. \& Sparke, L. S. 1986, Astrophys. J., 305, 175
\bibitem{larsson2012}
Larsson, S. et al., 2012, these proceedings
\bibitem{lott2012}
Lott, B., Escande, L., Larsson, S. \& Ballet, J., 2012, submitted to 
Astronomy \& Astrophysics, arXiv:1201.4851
\bibitem{maoznetzer1989}
Maoz, D., \& Netzer, H. 1989, MNRAS, 36, 21
\bibitem{maxmoerbeck2012}
Max-Moerbeck, W. et al., 2012, these proceedings
\bibitem{peterson1993}
Peterson, B. M. 1993, PASP, 105, 247
\bibitem{peterson1998}
Peterson, B. M. et al., 1998, PASP, 110, 660
\bibitem{scargle2011}
Scargle, J. D., 2011, Proceedings of the Workshop "Fermi meets Jansky - 
AGN in Radio and Gamma-Rays", Savolainen, T., Ros, E., Porcas, R.W. \& 
Zensus, J.A. (eds.), MPIfR, Bonn, June 21-23 2010
\bibitem{scargle2012}
Scargle, J. D., et al., 2012, in preparation.
\bibitem{smithvaughan2007}
Smith, R. \& Vaughan, S., 2007, MNRAS, 375, 1479
\bibitem{timmer1995}
Timmer, J., \& Koenig, M. 1995, Astronomy \& Astrophysics, 300, 707
\bibitem{wadehorne1988}
Wade, R.A. \& Horne, K. 1988, Astrophys. J., 324, 411
\bibitem{welsh1999}
Welsh W. F., 1999, PASP, 111, 1347
\bibitem{white1994}
White, R. J. \& Peterson, B. M., 1994, PASP, 106, 879
\end{thebibliography}

\end{document}